\documentclass[11pt, oneside]{article}

\usepackage[paper=a4paper, total={16cm,23cm}]{geometry}
\usepackage[T1]{fontenc}
\usepackage{setspace} 
\usepackage[hang]{footmisc} 
\usepackage{titling} 
\usepackage{tocloft} 
\usepackage{parskip} 

\usepackage{amsfonts}
\usepackage{amsmath}
\usepackage{amssymb}
\usepackage{mathrsfs, dsfont, amsthm}
\usepackage{physics}
\usepackage{mathtools}
\numberwithin{equation}{section} 

\usepackage{xcolor}
\definecolor{dark-red}{rgb}{0.50,0.12,0.12} 
\definecolor{mblue}{rgb}{0.30, 0.45, 0.70}
\definecolor{mred}{rgb}{0.70, 0.20, 0.20}
\definecolor{mgray}{rgb}{0.63, 0.63, 0.63}

\usepackage{hyperref} 
\hypersetup{colorlinks=true, allcolors=dark-red, linktoc=all}

\usepackage{cite}

\usepackage{tikz}
\usepackage{tikzlings}
\usetikzlibrary{calc, patterns, decorations, decorations.markings, shapes, positioning, intersections, quotes, angles}
\usepackage[subrefformat=parens]{subcaption}
\usepackage{pgfplots}
\pgfplotsset{compat=newest}
\newcommand{\mathdefault}[1][]{}

\def \cD {\mathcal{D}}
\def \cH {\mathcal{H}}
\def \cI {\mathcal{I}}
\def \cK {\mathcal{K}}

\def \bbR {\mathbb{R}}
\def \bbT {\mathbb{T}}
\def \bbZ {\mathbb{Z}}

\newcommand{\HH}{\mathrm{HH}}
\newcommand{\SL}{\mathrm{SL}}
\newcommand{\SO}{\mathrm{SO}}
\newcommand{\ep}{\mathrm{e}}
\newcommand{\ic}{\mathrm{i}}
\newcommand{\diff}{\mathrm{d}}
\newcommand{\dS}{\text{dS}}

\newcommand{\GHY}{\text{{GHY}}}

\newcommand{\Arg}[1]{\text{Arg}\,#1}

\newcommand{\defeq}{\mathrel{\rlap{\raisebox{0.3ex}{$\cdot$}}\raisebox{-0.3ex}{$\cdot$}}=}
\newcommand{\overbar}[1]{\mkern 1.3mu\overline{\mkern-1.3mu#1\mkern-1.3mu}\mkern 1.3mu}

\makeatletter
\newcommand{\subalign}[1]{
  \vcenter{
    \Let@ \restore@math@cr \default@tag
    \baselineskip\fontdimen10 \scriptfont\tw@
    \advance\baselineskip\fontdimen12 \scriptfont\tw@
    \lineskip\thr@@\fontdimen8 \scriptfont\thr@@
    \lineskiplimit\lineskip
    \ialign{\hfil$\m@th\scriptstyle##$&$\m@th\scriptstyle{}##$\hfil\crcr
      #1\crcr
    }
  }
}
\makeatother

\usepackage{mathtools}
\usepackage{comment}
\newcommand{\beq}{\begin{equation}}
\newcommand{\eeq}{\end{equation}}
\newcommand{\del}{\partial}

\begin{document}
\begin{titlingpage}
    \vspace*{3em}
    \onehalfspacing
    \begin{center}
        {\LARGE The spectrum of pure dS$_3$ gravity in the static patch}
    \end{center}
    \singlespacing
    \vspace*{2em}
    \begin{center}
        \textbf{
        Joydeep Chakravarty,$^1$
        Alexander Maloney,$^{1,2,3}$
        Keivan Namjou,$^1$
        and Simon F. Ross$^4$
        }
    \end{center}
    \vspace*{1em}
    \begin{center}
        \textsl{
        $^1$\ Department of Physics, McGill University \\
        Montr\'eal, QC H3A 2T8, Canada \\[\baselineskip]
        $^2$\ Department of Physics, Syracuse University \\
        Syracuse, NY, 13244, USA \\[\baselineskip]
        $^3$\ Institute for Quantum and Information Sciences, Syracuse University \\
        Syracuse, NY, 13244, USA \\[\baselineskip]
        $^4$\ Centre for Particle Theory, Department of Mathematical Sciences, Durham University \\
        South Road, Durham DH1 3LE, UK \\[\baselineskip]
        }
        \href{mailto:joydeep.chakravarty@mail.mcgill.ca}{\small joydeep.chakravarty@mail.mcgill.ca},
        \href{mailto:admalone@syr.edu}{\small admalone@syr.edu} \\
        \href{mailto:keivan.namjou@mail.mcgill.ca}{\small keivan.namjou@mail.mcgill.ca},
        \href{mailto:joydeep.chakravarty@mail.mcgill.ca}{\small s.f.ross@durham.ac.uk}
    \end{center}
    \vspace*{3em}
    \begin{abstract}
    We consider the quantum mechanical description of the de Sitter static patch in three-dimensional general relativity.  We consider a Lorentzian path integral that conjecturally computes the Fourier transform of the spectrum of the static patch Hamiltonian. We regulate a saddle point for this integral by a complex deformation that connects it to future infinity. Our computation is thus closely connected with the wave function of de Sitter gravity on a torus at future infinity.  Motivated by this, we identify an infinite number of saddle points that contribute to our Lorentzian path integral. Their sum gives a surprisingly simple result, which agrees with the expected features of the de Sitter static patch. For example, the thermal entropy, evaluated at the de Sitter temperature, agrees with the Bekenstein-Hawking formula. We also obtain a spectrum in the spin-zero sector, which is bounded, discrete, and has an integer degeneracy of states. It includes a dense spectrum of states, making both positive and negative contributions to the trace, arranged in such a way that negative contributions are invisible in the computation of any smooth observable. Nevertheless, several mysteries remain.
    \end{abstract}
\end{titlingpage}
\tableofcontents

\section{Introduction}
The quantum mechanical description of de Sitter space is perhaps one of the most important stepping stones in our understanding of quantum cosmology. Much like a black hole, de Sitter space has an event horizon that emits thermal radiation and is expected to have a Bekenstein-Hawking entropy \cite{Gibbons:1977mu}. However, the exact nature and interpretation of this entropy remain mysterious, in part because the horizon is observer-dependent.

One natural approach is to consider the thermodynamic interpretation of physics in the static patch, which is the causal region associated with a single observer. Various aspects of physics with a static patch observer have been previously studied in \cite{Anninos:2011af, Nakayama:2011qh, Chandrasekaran:2022cip, Loganayagam:2023pfb, Kolchmeyer:2024fly}. This contrasts with other approaches to the quantum mechanics of de Sitter space which focus on the wave function of the universe \cite{Hartle:1983ai}, which can be studied either on a timelike surface or near the asymptotic boundary $\cI^+$; in the latter case, this can be interpreted in the language of the dS/CFT correspondence as related to the partition function of some holographically dual conformal field theory \cite{Strominger:2001pn} (see also \cite{Witten:2001kn} for details of quantum gravity in de Sitter space). Various aspects of the correspondence have been solidified using lessons from holography in AdS \cite{Maldacena:2002vr, Castro:2012gc, Chakraborty:2023yed}. On a similar note, there have also been attempts to realize the correspondence by providing a microscopic description \cite{Anninos:2011ui, Anninos:2017eib, Collier:2025lux}.

In this paper, we will study the physics of the static patch of de Sitter space using Lorentzian path integral techniques. We will consider pure three-dimensional gravity, i.e., Einstein-Hilbert gravity, with a positive cosmological constant. The metric in the static patch is given by
\begin{equation}\label{staticp}
     \diff s^2 = - \qty( \ell^2-{r^2} ) \diff t^2 + \frac{\diff r^2}{1-\frac{r^2}{\ell^2}} + r^2 \diff \phi^2. 
\end{equation}
An observer sitting at $r=0$ in these coordinates will see an event horizon at $r=\ell$. In order to understand the quantum mechanics of this observer, our goal is to compute the following observable:
\begin{equation}\label{eq:trace}
   Z(T) \defeq \Tr (\ep^{- \ic HT}),
\end{equation}
where $H$ is the bulk Hamiltonian generating translation in $t$ and the trace is over the  Hilbert space of quantum gravity in the static patch. Of course, given the subtle nature of de Sitter quantum gravity, it is not a priori clear that such a Hilbert space exists or what properties it should have. Does it have a positive norm?  Is the spectrum of $H$ discrete and bounded?  Fortunately, in three dimensions we have a significant advantage in trying to answer these questions: we can, with certain assumptions, simply compute $Z(T)$ exactly.
 
Our starting point is the observation that the trace $Z(T)$ is formally given by a Lorentzian path integral, which has a saddle-point given by the above geometry with the timelike identification $t \sim t + T$. Although a periodic identification in Lorentzian time may appear problematic, similar Lorentzian periodic identifications have been considered for black hole solutions, where they give rise to the so-called double cone geometries \cite{Saad:2018bqo}. In AdS, these geometries have two boundaries, coming from the two asymptotic regions of the black hole, and the resulting solution is interpreted as a saddle-point contribution to the spectral form factor $$\overbar{\Tr(\ep^{\ic H T}) \Tr(\ep^{-\ic HT})}$$ in the dual CFT \cite{Saad:2018bqo}. Our goal is to apply a similar logic to understand the trace (\ref{eq:trace}) in de Sitter space.

In both the black hole and de Sitter cases, the periodic identification in $t$ has a fixed point at the horizon, which renders the solution singular. In \cite{Saad:2018bqo}, this singularity was resolved by complexifying the geometry in such a way that it remains a solution to the (complexified) equations of motion. 
This is somewhat similar to more familiar Wick rotations and was further clarified in \cite{Chen:2023hra}, who showed that this procedure is equivalent to replacing $H$ by a modified boost Hamiltonian $\tilde{K}$. In this way, the double cone geometry is interpreted as a saddle-point contribution to the bulk Hilbert space trace $\Tr_{\text{bulk}}(\ep^{-\ic \tilde{K} T})$. This was recently generalized in \cite{Chakravarty:2024bna}, where other complex deformations resolving the fixed point singularity provide contributions to the amplitude for cosmological solutions in AdS.

\subsection{Methods and results}
In our study of de Sitter space, similar to the double cone geometry, we consider complex deformations of the periodically identified static patch \eqref{staticp}. As with the double-cone geometries, these are obtained by complexifying the radial coordinate $r$ in the static patch metric, so they can similarly be interpreted in terms of a modification of the static patch Hamiltonian. Unlike in the double cone, these deformations do not satisfy the Kontsevich-Segal condition \cite{Kontsevich:2021dmb}, which was argued in \cite{Witten:2021nzp} to be a natural condition to impose on complex contributions to the gravitational path integral. This condition is necessary in order to render well-defined the perturbative fluctuations of an arbitrary field in a complex background. In the present case, since three-dimensional general relativity has no local degrees of freedom, it is not so clear that this condition is necessary.  Indeed, we will discuss below the computation of loop corrections around each of our saddle points.

We will discover a particular class of complex geometries which can be used to resolve the singularity at the cosmological horizon. In particular, we will find a deformation that connects the static patch to the asymptotic future boundary, relating $\Tr (\ep^{- \ic HT})$ in the static patch to the wavefunction in the torus cosmology obtained by periodically identifying $t$ in the region to the future of the cosmological horizon  (see the complex radial contour in Fig. \ref{fig:sp-to-cosmo}). This means that our trace $Z(T)$ has a second interpretation in terms of the wave function of the universe for space-times of the form $\bbT^2 \times \mathbb{R}_{\text{time}}$, where $\bbT^2$ is a torus. As one approaches $\cI^+$, the parameter $T$ has a natural interpretation in terms of the conformal structure $\tau$ of a torus at future infinity. This provides a connection between our work and closely related considerations in the context of the dS/CFT correspondence \cite{Bousso:2001mw, Balasubramanian:2001nb}.

One key feature of our work is that there is, in fact, an infinite number of ways of resolving the cosmological singularity. Indeed, in \cite{Castro:2012gc} (following \cite{Maldacena:2002vr}), a proposal for the wavefunction of the universe on a torus in pure three-dimensional quantum gravity with a positive cosmological constant was obtained by analytic continuation from the case with negative cosmological constant: 
\begin{equation}\label{eq:psi}
    \Psi_\HH(\tau) = \Psi_{0,1}(\tau) + \sum_{c,d} \Psi_{c,d}( \tau), 
\end{equation}
where $\tau$ is the complex modular parameter for the torus, and the sum is over relatively prime $c,d$ with $c>0$, corresponding to the non-trivial part of the modular group acting on the boundary torus. This expression is obtained by analytic continuation from the AdS case \cite{Maloney:2007ud}, where the Poincar\'e sum over $c,d$ can be interpreted as summing over the different ways of filling in the boundary torus with a Euclidean AdS$_3$ bulk, where different boundary cycles correspond to the contractible cycle in the bulk. In \cite{Castro:2012gc}, this was interpreted as mapping over to a sum over Lorentzian bulks consisting of a future de Sitter cosmological region with a past singularity where one of the cycles collapses at the cosmological singularity. 

In our present work, we can give an entirely different interpretation to these results. We will argue that each of the saddle points labeled by $(c,d)$ in this expression gives a different contribution to the trace $Z(T)$ defined in equation (\ref{eq:trace}). In terms of the static patch, each represents a different way of resolving the singularity at the cosmological horizon (by connecting onto $\cI^+$ in a different way), so that the full result is given by a sum over these Lorentzian saddle points. Indeed, we will be able to compute the Hilbert space trace $\Tr (\ep^{- \ic H T+\ic \theta J})$ in terms of a Poincar\'e series related to a torus with modular parameter $\tau= \theta + \ic T$.

In fact, the wave function on $\bbT^2 \times \mathbb{R}_{\text{time}}$ was recently considered in \cite{Godet:2024ich}, where it was shown that the wavefunction \eqref{eq:psi} can be rewritten as a sum over exponentials $\ep^{- \ic E \Im \tau}$. Our relation to the static patch trace offers a natural interpretation of this sum in terms of the spectrum of states in the static patch Hilbert space. So one of our results is a re-interpretation of \cite{Godet:2024ich} in terms of the Hilbert space of de Sitter quantum gravity in the static patch, giving insight into the states underlying the de Sitter entropy. We will comment on some subtleties in the derivation of the formula in \cite{Godet:2024ich}, and discuss the inclusion of contributions from Maas cusp forms. 

Our final result is an answer for $\Tr (\ep^{- \ic H T + \ic \theta J})$, which can be interpreted in terms of the spectrum of de Sitter quantum gravity in the static patch. Focusing on the $J=0$ sector, we find several unusual (and compelling) features:
\begin{enumerate}
    \item The spectrum of $H$ is bounded below, in the sense that the real parts of the energies are bounded.
    \item The spectral degeneracies are integers.
    \item The spectrum is discrete but dense.
    \item The spectrum contains states that contribute negatively to the trace.
    \item The semi-classical part of the thermal entropy -- evaluated at the de Sitter temperature---agrees with the expected result for the de Sitter entropy $S_{\dS} = \frac{\pi \ell}{2G} + \cdots$.
\end{enumerate}
The first two features listed above are just what one might expect of a normal quantum mechanical system, and it is particularly encouraging that integer spectral degeneracies are derived directly from a gravitational path integral.
The second two features are not, however, and suggest that we are not dealing with a standard Hilbert space. Nevertheless, we will argue that they are exactly what we need in order to explain the physics of the de Sitter static patch.  First, we note that because an observer living in de Sitter will have a finite lifetime (due to the finite temperature), no observer will be able to resolve individual states in the dense spectrum.  Moreover, the positive and negative degeneracies in the spectrum are arranged in such a way that they effectively cancel out in the computation of any smooth observable. Indeed, this cancellation plays an important role in recovering the finite de Sitter entropy, $S_{\dS}$, described in the fifth point.

Our computation of the de Sitter entropy should be compared to the more familiar approach, where one computes the thermal partition function $\Tr (\ep^{-\beta H})$ using a Euclidean path integral. In that case, the de Sitter entropy is obtained from a saddle point described by the analytically continued geometry $t \to \ic \tau$, with $\tau$ periodic, $\tau \sim \tau + \beta$ \cite{Gibbons:1976ue}. In our case, we are obtaining it from a Lorentzian saddle point, with time periodically identified ($t\sim t+T$), and effectively analytically continuing $T$ to imaginary values.
Either quantity encodes the spectrum of states of the de Sitter static patch. Our approach differs in that we are able to identify an infinite number of subleading additional saddles, which allow one to completely reconstruct the spectrum.
\footnote{Matching the de Sitter entropy beyond the semi-classical limit is subtle, however, due to divergences in the one-loop determinants. We will comment on this below.}

This way of understanding the de Sitter entropy also allows us to more properly understand an old observation of \cite{Bousso:2001mw, Balasubramanian:2001nb}, where it was noted that applying an analytically continued Cardy formula in the context of dS/CFT could reproduce the de Sitter entropy. Here, we can justify this formula as a calculation of the dominant saddle-point contribution to the analytically continued trace.\footnote{More precisely, the computations in these papers worked because they include two factors of $\ic$ which happen to cancel out.  First, these papers considered the torus partition function of the conjectured CFT dual of dS$_3$, but were normalizing the stress tensor with an unconventional minus sign; with the correct normalization, the central charge is imaginary in the semi-classical limit.  Second, they were considering this Euclidean CFT on a torus with temperature given by the Hawking temperature $\beta=\beta_{\dS}$.  However, the analytic continuation to the static patch -- as given by the complex saddle described above -- identifies $\beta$ with $T$, the periodicity in Lorentzian static patch time.  So the correct procedure is to set $\beta = \ic \beta_{\dS}$.  With these two modifications, these earlier calculations can be understood as the semi-classical part of the entropy computation presented in this paper.} 

The organization of the rest of the paper is as follows. In the next section, we briefly review the de Sitter solutions in pure three-dimensional gravity in both the static patch and torus cosmology. We then consider complex contours that connect the two and discuss the issues with satisfying the KSW condition. Section \ref{rel} contains our main proposal, describing the sum over Lorentzian geometries which we are computing and how the trace over the static patch is related to the wavefunction \eqref{eq:psi}. We check this proposal by showing that the action for the complex contours matches the exponential part of the terms in the Poincar\'e sum and discuss the calculation of the de Sitter entropy from this perspective. Then in section \ref{spectral} we relate the trace over the static patch to the calculation of \cite{Godet:2024ich}, and discuss some features and extensions of that calculation.

\section{Complex contours in three-dimensional de Sitter}\label{contour}
We are interested in understanding the de Sitter static patch with a periodic identification in Lorentzian time. We consider pure gravity in three dimensions with a positive cosmological constant, with action
\begin{equation}\label{act}
S = \frac{1}{16 \pi G} \int_\mathcal{M} \diff^3x \, \sqrt{-g} (R - 2\Lambda) + \frac{\epsilon}{8\pi G} \int_{\del \mathcal{M}} \diff^2 x \, \sqrt{h} K + S_\mathrm{ct},
\end{equation}
where the second term is the GHY boundary term,\footnote{The parameter $\epsilon$ takes the value $\epsilon=1$ when the normal to the boundary $\del \mathcal{M}$ is spacelike, and $\epsilon=-1$ when the normal to the boundary is timelike.} while the final term is a boundary counter-term. In three dimensions, the cosmological constant is related to the de Sitter length scale $\ell$ as $\Lambda = 1/\ell^2$.

This action has a de Sitter solution, which in the static patch coordinates is 
\begin{equation}
    \diff s^2 = - \qty( \ell^2 - {r^2} ) \diff t^2 + \frac{\diff r^2}{1 - \frac{r^2}{\ell^2}} + r^2 \diff \phi^2,
\end{equation}
with $r \in (0,\ell)$. A static observer living at $r =0$ sees a cosmological horizon at $r = \ell$.\footnote{Although we have used the word ``observer'' here, we are not adding physical degrees of freedom associated with the observer, as in e.g. \cite{Chandrasekaran:2022cip}.  This would be interesting to consider in the future, and may provide insight into the interpretation of our results.} This is analogous to the black hole exterior with an important distinction: the horizon is not a global feature of the spacetime, but instead depends on the location of the observer. Note that we have chosen the coordinates $t$ and $\phi$ as dimensionless, and hence the de Sitter inverse-temperature is given by $\beta_{\dS} =2\pi$.

We will also find it useful to consider the "future" region with $r > \ell$ and the metric
\begin{equation}
    \diff s^2 = \qty(r^2 - {\ell^2}) \diff t^2 - \frac{\diff r^2}{\frac{r^2}{\ell^2} - 1} + r^2 \diff \phi^2.
\end{equation}
The metric is the same, but now $r$ plays the role of a time coordinate, with the horizon in the past at $r=\ell$, and the asymptotic boundary ${\cI}^+$ at $r\to\infty$.  This is somewhat analogous to the "interior" patch of a black hole geometry.

We are interested in studying the trace
\begin{equation}\label{trsp}
Z(T) = \Tr_\cH {\ep^{-\ic H T}},
\end{equation}
over the static patch Hilbert space $\cH$. Here $H$ is the Hamiltonian which generates static patch time translations, realized geometrically as $H=-\ic\partial_t$.  This can be generalized to consider a trace
\begin{equation}
Z(T, \theta) = \Tr_\cH {\ep^{-\ic H T-\ic \theta J}},
\end{equation}
where the angular momentum $J=-\ic \partial_\theta$ generates rotations, and $\theta$ is a corresponding angular potential.  Of course, since $J\in\mathbb{Z}$ is quantized $\theta\sim\theta+2\pi$ is periodic.

In general, we do not have a direct way of constructing the static patch Hilbert space $\cH$, nor do we even necessarily know that it possesses all of the usual features of a quantum mechanical Hilbert space.  Instead, our approach will be to compute $Z(T,\theta)$ from a Lorentzian path integral, and from this infer what properties $\cH$ must have.
 
Our starting observation is that one natural saddle-point which should contribute to $Z(T,\theta)$ is the static patch geometry itself, with the following periodic identification implemented by the Hilbert space trace:\footnote{A periodic identification along the $t$ circle in the static patch was also recently discussed in \cite{Banihashemi:2024weu}.}
\begin{equation}\label{identifications}
\phi \sim \phi + 2\pi+\theta, \qquad t \sim t + T. 
\end{equation}
For convenience, let us just consider the case $\theta=0$.
In this case, the region $r > \ell$ gives us the torus cosmology, studied for example in \cite{Maloney:2007ud, Anninos:2011af, Godet:2024ich}. In the static patch region, the surface of fixed $r$ is a Lorentzian torus, and the geometry has closed time-like curves. The $\phi$ circle closes off smoothly at the observer's location at $r=0$, while the surface $r=\ell$ has a Milne-type singularity where the identification in $t$ has fixed points. This is analogous to the singularity in the double cone geometry at the black hole horizon \cite{Saad:2018bqo}.

\subsection{Complex contours}
The saddle point geometry described above is singular, and in particular has a location ($r=\ell$) where the timelike circle shrinks to zero size. We would like to resolve this singularity somehow.  Here we will follow \cite{Saad:2018bqo}, and complexify the geometry slightly.  This complexification can be understood as obtaining another (slightly complex) solution to the equations of motion. Alternatively, following \cite{Maldacena:2024uhs}, it can also be understood as imposing a condition on the wave functions which are allowed to propagate in our space-time. When complexifying our geometry, we would like to maintain the time-translation and angular-translation symmetries.  Effectively, this means that we should complexify our radial coordinate $r$.

We will begin by introducing an isotropic coordinate $r = \ell \sin u$, in terms of which the metric is
\begin{equation}\label{dsmetric}
    \frac{\diff s^2}{\ell^2} = -\cos^2 u  \, \diff t^2 +  \diff u^2 +  \sin^2 u \, \diff \phi^2,
\end{equation}
Here the static patch is the region $u \in \qty(0, \frac{\pi}{2})$, with our observer at $u=0$ and horizon at $u = \frac{\pi}{2}$. 
Similarly, taking
\begin{equation}\label{futuretr}
u = \frac{\pi}{2} + \ic \xi, 
\end{equation}
the region $\xi \in \mathbb{R}^+$ describes the future region where we have the torus cosmology. The metric is
\begin{equation}\label{futuremetric}
\frac{\diff s^2}{\ell^2} = - \diff \xi^2 + \sinh^2 \xi  \, \diff t^2  +  \cosh^2 \xi \, \diff \phi^2,
\end{equation}
spacelike infinity $\mathcal{I}^+$ is at $\xi \to \infty$.

To construct a complex solution which preserves both time-translation and rotation symmetry, we can now just consider a contour in the complex $u$ plane.  Specifically, taking $u=u(\rho) \in \mathbb{C}$, for some real coordinate $\rho$, the metric is 
\begin{equation}\label{dsmetricc}
    \frac{\diff s^2}{\ell^2} = -\cos^2 u(\rho)  \, \diff t^2 +  u'(\rho)^2 \diff \rho^2 +  \sin^2 u(\rho) \, \diff \phi^2,
\end{equation}
which is a complex solution to the equations of motion.

In this way, it is simple to resolve the singularity at $u=\pi/2$---we simply consider a contour that goes around this point in the complex $u$ plane.  Indeed, based on our experience in AdS, it is natural to consider a contour that connects the static patch to the torus cosmology.  In the case of the BTZ black hole, an analogous contour connects the cone geometry to a cosmology inside the horizon \cite{Chakravarty:2024bna}.

In studying complex solutions, one important consideration is whether the geometry satisfies the Kontsevich-Segal-Witten (KSW) criterion \cite{Kontsevich:2021dmb, Witten:2021nzp}. The condition states that the eigenvalues of the metric should satisfy 
\begin{equation}\label{ksw}
\sum_{\mu=1}^3 \abs{\Arg{\lambda_\mu}} <\pi . 
\end{equation}
In physical terms, this guarantees that the perturbative fluctuations of an arbitrary spin field are well behaved at the one-loop level.
In the present case, we are considering pure gravity, rather than a generic field theory, so the KSW condition is stronger than what we need.  Nevertheless, it is interesting to ask whether it is possible to find contours that resolve the singularity at $u=\pi/2$ and obey KSW.

To begin, consider the metric \eqref{dsmetricc} with $u(\rho)= \rho + \ic \xi(\rho)$. We have
\begin{equation}
\sum_{\mu=1}^3 \abs{\Arg{\lambda_\mu}}  = 2 \arctan \abs{\frac{\diff \xi}{\diff \rho}} + 2 \arctan(\cot \rho \tanh \xi) + 2 \arctan(\cot \rho \coth \xi). 
\end{equation}
The last two terms together are precisely equal to $\pi$ when $\rho= \frac{\pi}{4}$. For $\rho < \frac{\pi}{4}$, they are in general greater than $\pi$, and only equal $\pi$ on the real axis at $\xi=0$. Thus, near the origin of the static patch, there is no complex contour that satisfies the KSW condition. The real Lorentzian metric on the static patch saturates the bound, but it is impossible to do better simply by considering complex $u$. Furthermore, if we start exactly on the real axis, we cannot deviate from it at larger $\rho$ while satisfying the bound, as the last two terms remain equal to $\pi$ all along the real axis, and lifting the contour up off the axis requires us to make the first term non-zero, violating the bound \eqref{ksw}. Thus, while there are contours in the region $\rho > \frac{\pi}{4}$ which locally satisfy the KSW condition, we cannot access them starting precisely on the real axis. 

This seems to be a general obstruction to the construction of complexifications of the de Sitter static patch that obey the KSW condition.  In the present case, where we are considering pure gravity, we know that the KSW condition is more than we need.  So we will proceed with our complex contour construction.  However, it is worth noting that this may indicate 
more general issues with the stability of perturbative fields in de Sitter. However, we note that the obstruction to satisfying KSW is associated with the region near the origin of the static patch, separated from the singularity at $u = \frac{\pi}{2}$, which we want to resolve. It is also worth noting that the use of the KSW bound in restricting the path integral is only conjectured, and there are other cases where apparently interesting contours violate it, including the bra-ket wormholes \cite{Chen:2020tes, Fumagalli:2024msi} and the double cone geometries in higher-dimensional AdS black holes (see the appendix of \cite{Chakravarty:2024bna}). 

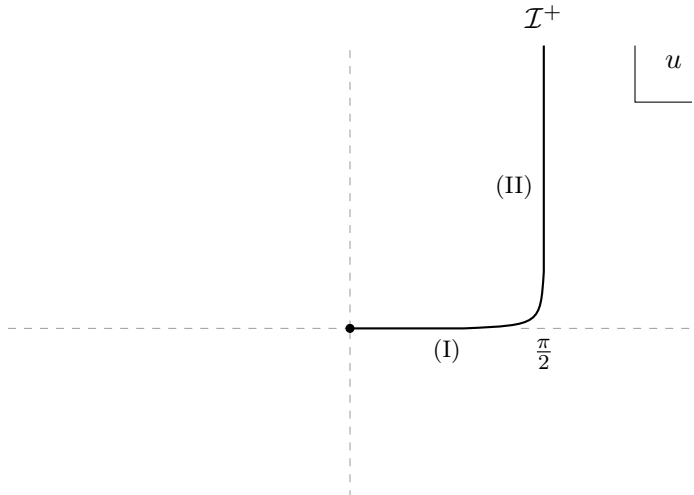
\begin{figure}
\centering
\begin{tikzpicture}[scale=1.5]
    \draw[dashed,mgray] (-3,0)--(0,0) (1.5,0)--(3,0) (0,-1.5)--(0,0) (0,0)--(0,2.5);
    \draw (2.5,2.5)--(2.5,2)--(3,2) (3,2.5)node[anchor=north east]{$u$};
    \draw[thick] (0,0)--(.85,0)node[anchor=north]{\footnotesize (I)}--(1,0)..controls(1.67,.03)..(1.7,.5)--(1.7,1.25)node[anchor=east]{\footnotesize (II)}--(1.7,2.5);
    \draw (1.7,0)node[anchor=north]{$\frac{\pi}{2}$};
    \filldraw (0,0) circle (1pt);
    \draw (1.7,2.95)node[anchor=north]{$\cI^+$};
\end{tikzpicture}
\caption{The contour connecting the static patch (in dot) to the spacelike infinity $\cI^+$ in the complex $u$-plane.}\label{fig:sp-to-cosmo}
\end{figure}

To proceed, we must consider a geometry that resolves the singularity in the periodically identified static patch at $u=\pi/2$.  We will do so by considering  
contours that start on or near the real $u$ axis until $u$ gets close to $\frac{\pi}{2}$, where they smoothly rotate to a near-vertical contour, approaching the cosmology $u = \frac{\pi}{2} + \ic \xi$ as $\xi \to \infty$. Such contours connect the identified static patch to the torus cosmology, as shown in figure \ref{fig:sp-to-cosmo}.

Therefore, this complex contour provides a smooth saddle-point contribution to a version of the trace over the static patch.  Specifically, the deformation near the singularity corresponds to a deformation in the definition of the static patch Hamiltonian, as in the discussion of the double cone in \cite{Chen:2023hra}.

\section{Relation to the de Sitter wave function}\label{rel}
We propose that if we resolve the singularity at the cosmological horizon by the complex deformation above, this relates the trace over the static patch to the wavefunction of de Sitter space defined on the future spacelike boundary $\cI^+$. 

A proposal for this wavefunction was given in \cite{Castro:2012gc} (based on \cite{Maldacena:2002vr}) by considering the MWK partition function in AdS and analytically continuing it to de Sitter space. The MWK partition function, which describes three-dimensional pure gravity, takes the form of a Poincar\'e sum:
\begin{equation}\label{mwk}
    Z_{\rm MWK}(\tau) =  Z_{0,1}(\tau) + \sum_{c,d} Z_{c,d}(\tau),
\end{equation}
where $\tau= x + \ic y$ is the modular parameter of the boundary torus, $Z_{0,1}(\tau)$ denotes the thermal AdS configuration, and the other terms are its images under the non-trivial part of the modular group acting on the boundary torus, $\SL(2,\bbZ)/\bbZ$. The sum is over coprime integers $c$ and $d$ with $c > 0$. 
\footnote{For details on notation and the MWK partition function, see appendix \ref{mwkpf}.} The analytic continuation from Euclidean AdS to Lorentzian dS takes us from the MWK partition function to the late-time Hartle-Hawking wavefunction. This takes the following form:
\begin{equation}\label{HH1}
    \Psi_\HH(\tau) =  \Psi_{0,1}(\tau) + \sum_{c,d} \Psi_{c,d}( \tau), 
\end{equation}
where the component $\Psi_{0,1}(\tau)$ obtained from the continuation of $Z_{0,1}(\tau)$ is\footnote{The value of the central charge will be discussed in the next section.}
\begin{equation}\label{HH2}
\Psi_{0,1}(\tau) = \frac{1}{\abs{\eta(\tau)}^2} \abs{q}^{-(c_\dS-1)/12} \abs{1-q}^2, \qquad c_\dS = 13 + \ic \frac{3\ell}{2G},
\end{equation}
with $q = \ep^{2 \pi \ic \tau}$, and the other terms in the Poincar\'e sum are\footnote{The solution for $a,b$ given $c,d$ is ambiguous up to shifts by $a \to a + m c$, $b \to b + m d$, but since $\Psi_{0,1}$ is invariant under such shifts the expression for $\Psi_{c,d}(\tau)$ is unambiguous.}
\begin{equation}
\Psi_{c,d}(\tau) = \Psi_{0,1}(\gamma \tau), \qquad \text{where} \qquad \gamma \tau = \frac{a \tau+b}{c\tau + d}, \quad ad-bc =1.
\end{equation}

In considering the modular images, it is sometimes convenient to take an overall factor of $\sqrt{\Im \tau} \abs{\eta(\tau)}^2$ outside in the sum \eqref{HH1}, as this factor is modular invariant; thus we can also write \eqref{HH1} as 
\begin{equation} 
    \Psi_\HH(\tau) =  \frac{1}{\sqrt{\Im \tau} \abs{\eta(\tau)}^2} \qty( \psi_{0,1}(\tau) + \sum_{c,d} \psi_{c,d}( \tau) ), 
\end{equation}
where 
\begin{equation}
\psi_{0,1}(\tau) = \sqrt{\Im \tau} \abs{q}^{-(c_\dS-1)/12} \abs{1-q}^2, \qquad c_\dS = 13 + \ic \frac{3\ell}{2G},
\end{equation}
and $\psi_{c,d}(\tau) = \psi_{0,1}(\gamma \tau)$.

In AdS, the partition function \eqref{mwk} corresponds to a sum over bulk geometries that fill in the boundary torus. The deformed contour which we constructed in the previous section as an approximation to the trace over the static patch, extends up to $\cI^+$, so it should also contribute to the calculation of this wavefunction. In \cite{Castro:2012gc}, the terms in the Poincar\'e sum were interpreted as corresponding to different semiclassical saddles filling in the boundary torus, as in the AdS case. There, the semiclassical saddles considered were taken to be the real Lorentzian cosmology \eqref{futuremetric} and its images under the modular transformation. These semiclassical geometries have a singularity at $\xi=0$, where the $t$ circle degenerates---this is the same singularity that we discussed ``from the other side'' in our previous discussion of the static patch. The resolution of this singularity was not previously addressed. We now want to propose that the singularity is resolved by considering instead the complex contour introduced above, which closes off smoothly at the origin of the static patch. 

To correspond to our previous discussion of the trace over the static patch \eqref{trsp}, that is, to the identifications \eqref{identifications}, we should consider a rectangular boundary torus, with $\tau = \ic \frac{T}{2\pi}$. This could be easily generalized to consider general tori with $\tau = \theta + \ic \frac{T}{2\pi}$, by considering $\Tr (\ep^{- \ic HT + \ic \theta J})$, but we will continue to restrict to the case with $\theta=0$ for simplicity.

\subsection{Classical Einstein-Hilbert action}\label{cact}
In the AdS case, the classical action of thermal AdS produces the leading exponential behavior of $Z_{0,1}$. The real Lorentzian cosmology is obtained by analytic continuation of thermal AdS, so its classical action should similarly reproduce the leading behavior of $\Psi_{0,1}$. One might worry that by deforming the contour we could change the classical action, spoiling this agreement. Here we see that this is not the case. 

The bulk action of three-dimensional pure gravity is given in \eqref{act}. Substituting the dS metric in the action, we find that $R-2\Lambda = 4 \Lambda = 4/ \ell^2$. Using these values in \eqref{act}, the action becomes
\begin{equation}
S = \frac{1}{4 \pi G \ell^2} \int \diff^3x \, \sqrt{-g} + \frac{\epsilon}{8\pi G} \int \diff^2 x \, \sqrt{h} K - \frac{\epsilon}{8\pi G} \int \diff^2 x \, \sqrt{h},
\end{equation}
where we have also inserted the counterterm action $S_\mathrm{ct}$ in the last term \cite{Strominger:2001pn}. 

\paragraph{Future cosmology:} We first look at the future cosmology region \eqref{futuremetric}. The volume integral gives
\begin{equation}
    \int \diff^3x \, \sqrt{-g} = \ell^3 \Delta t \, \Delta \phi \int_0^{\xi_c} \diff \xi \, \sinh \xi \cosh \xi = \frac{\Delta t \, \Delta \phi}{4} (\cosh 2 \xi_c - 1),
\end{equation}
where $\xi_c$ denotes the cutoff at large $\xi$, i.e. at late time. The extrinsic curvature $K =  \ell^{-1}(\tanh \xi + \coth \xi)$, and $\sqrt{h} = \frac{\ell^2}{2} \sinh 2 \xi$. In the future cosmology region, since the normal to the spacelike boundary is timelike, we have $\epsilon = -1$. This implies that the GHY integrand takes the form $K \sqrt{h} = \ell \cosh 2 \xi$, so
\begin{equation}
S_\GHY = -\frac{\ell \Delta t \, \Delta \phi}{8\pi G} \qty(\cosh 2 \xi_c -1),
\end{equation}
where we are including contributions both at $\xi = \xi_c$ and at $\xi = 0$. Since the spacetime is not regular at $\xi=0$, we need to introduce a boundary there, and the GHY boundary term on this boundary gives a finite contribution to the action. As expected, the divergent terms in the volume integral and in the GHY term cancel against the divergent term coming from the counterterm action. As a result, we are only left with the finite terms coming from the Einstein-Hilbert volume term and  from the GHY action, 
\begin{equation}
S =  - \frac{\ell \Delta t \, \Delta \phi}{16 \pi G} + \frac{\ell \Delta t \, \Delta \phi}{8 \pi G}   =  \frac{\ell T}{8G}.
\end{equation}
Thus this Lorentzian saddle should contribute a term going as $\ep^{\ic S} = \ep^{\ic \frac{\ell T}{8G}}$. With $q = \ep^{2\pi \ic \tau} = \ep^{-T}$, this matches the exponential factor in $\Psi_{0,1}(\tau)$ in \eqref{HH1}. 

\paragraph{Along the radial contour:} We now extend the above analysis to our complexified radial contour. Recall that the contour starts at the static patch origin $u=0$ and runs along the static patch to the vicinity of $u = \frac{\pi}{2}$ and is smoothly glued to $u = \frac{\pi}{2} + \ic \xi$, and runs up to the future cosmology region, i.e. $\xi \to \infty$.

Let us now consider the volume contribution to the Einstein-Hilbert action. Taking the integrand in the volume as a holomorphic function of $u$ gives:
\begin{equation}
\begin{split}
    \int \diff^3x \sqrt{-g} &= \frac{\ell^3 \Delta t \, \Delta \phi}{2} \qty(\int_0^{\frac{\pi}{2}} \diff u \, \sin 2u  + \ic \int_0^{\xi_c} \diff \xi \, \sin (\pi + 2 \ic \xi)) \\
    &= \frac{\ell^3 \Delta t \, \Delta \phi}{2} \qty(\int_0^{\frac{\pi}{2}} \diff u \, \sin 2u  + \int_0^{\xi_c} \diff \xi \, \sinh (2 \xi)) \\
    &= \frac{\ell^3 \Delta t \, \Delta \phi}{4} \qty(2 + (\cosh 2 \xi_c - 1)).
\end{split} 
\end{equation}
This is essentially the sum of the contributions from the static patch region and from the future cosmology. Next, the GHY integrand over our new contour is supported only at one end, i.e., at the future cosmology, as the contour is smoothly capped off in the static patch end. Thus, we obtain the following contribution to the GHY action:
\begin{equation}
S_\GHY = -\frac{\ell \Delta t \, \Delta \phi}{8\pi G} \qty( \cosh 2 \xi_c).
\end{equation}
The counterterm action contribution is unchanged from the previous calculation. All the divergent pieces in the action cancel with each other as before, and we are left with the finite part of the total action,
\begin{equation}
S = \frac{\ell \Delta t \, \Delta \phi}{16 \pi G} = \frac{\ell T}{8 G}.
\end{equation}
Thus, the action of the complex contour agrees with the action of the previously considered real Lorentzian saddle. Replacing the latter by the former does not affect the match with the wavefunction, hence, we can think of this as a resolution of the past singularity from this perspective.

\subsection{Main claim: dS wavefunction as trace over the static patch spectrum}
We have argued that the complex semiclassical saddle contributing to the trace over the static patch Hilbert space also corresponds to the $\Psi_{0,1}$ part of the wavefunction (\ref{eq:psi}). Our main hypothesis, motivated by this, is to identify the full wavefunction $\Psi_\HH$ with $\Tr_{\mathcal H}(\ep^{-\ic HT})$.\footnote{For a rectangular torus; for general $\tau$, this corresponds to $\Tr_{\mathcal H}(\ep^{-\ic HT + \ic \theta J})$, as noted before.} That is, we take the expression \eqref{eq:psi} to \emph{define} what we mean at a microscopic level by the trace over the static patch Hilbert space. 

This is a surprising claim, as the full wavefunction involves a Poincar\'e sum. If we identify $\Psi_{0,1}$ with the complex contour where the $\phi$ circle shrinks smoothly at the origin of the static patch, the other contributions $\Psi_{c,d}$ correspond to $\SL(2, \mathbb{Z})$ images of this contour, where the cycle which closes off smoothly is identified with a linear combination of the $t$ and $\phi$ cycles in the boundary torus. For fixed values of the boundary modular parameter $\tau$, this will give different values for the period of the orthogonal cycle. From the boundary wavefunction perspective, it is natural to include this modular sum, but in the calculation of the trace over the static patch Hilbert space, it is not obvious what the role of these other saddles is. 

We will provide evidence for our hypothesis first by noting in the next subsection that we can reproduce the de Sitter entropy from the $\Psi_{1,0}$ contribution in a suitable analytic continuation. Then, in the following section, we will see that the formulas for the full wavefunction given in \cite{Godet:2024ich} have features that seem natural from the perspective of the trace over the static patch Hilbert space.

\subsection{The de Sitter entropy}
To obtain the de Sitter entropy, we should analytically continue $\Tr(\ep^{-\ic HT})$ to $T = \ic \beta$, to calculate the thermal trace $Z(\beta) = \Tr(\ep^{-\beta H})$, and calculate 
\begin{equation}\label{entropy}
    S_{\dS} = \qty(1 - \beta \frac{\partial}{\partial \beta} ) \ln Z, 
\end{equation}
at the de Sitter temperature $\beta = \beta_{\dS} = 2\pi$. With our identification of $\Tr(\ep^{-\ic HT})$ with $\Psi_\HH(\tau)$, we want to consider the expression for $\Psi_\HH(\tau)$ for a rectangular torus $\tau = \ic \frac{T}{2\pi}$ as a function of $T$ and analytically continue to imaginary $T = \ic \beta$. 

For a rectangular torus, the modular images have 
\begin{equation}
    \Im \gamma \tau = \frac{ \Im \tau}{d^2 + c^2 (\Im \tau)^2}, \qquad \Re \gamma \tau = \frac{a}{c} - \frac{d/c}{d^2 + c^2 (\Im \tau)^2},
\end{equation}
and 
\begin{equation}
\begin{split}\label{psicdbeta}
    \psi_{c,d}(\tau) = \psi_{0,1}(\gamma \tau) &= \sqrt{ \Im \gamma \tau} \, \ep^{\frac{(c_{\mathrm{dS}}-13)}{12} 2\pi \Im \gamma \tau} (\cosh 2\pi \, \Im \gamma \tau  - \cos 2\pi \, \Re \gamma \tau)
    \\ &= \sqrt{ \Im \gamma \tau} \, \ep^{\frac{\ic \ell}{8G} 2\pi \Im\gamma \tau} (\cosh 2\pi \, \Im \gamma \tau  - \cos 2\pi \, \Re \gamma \tau).
    \end{split}
\end{equation}

When we analytically continue $\Im \tau$ to imaginary values, $\Im \gamma \tau$ becomes imaginary, while $\Re \gamma \tau$ stays real,
\begin{equation}
\Im \gamma \tau  = \ic \frac{2\pi \beta}{4\pi^2 d^2 - c^2 \beta^2}, \qquad \Re \gamma \tau = \frac{a}{c} - \frac{4\pi^2 d / c}{4\pi^2 d^2 - c^2 \beta^2}.
\end{equation}
As a result, the exponential phase in $\psi_{c,d}$ becomes a real exponential. For $\psi_{0,1}$, the exponential is negative, 
\begin{equation}\label{c0}
\psi_{0,1}(\tau) =  \ep^{-\frac{\ell \beta}{8G}} (\cos \beta - 1),
\end{equation}
while for the S-transformed saddle $\psi_{1,0}$, 
\begin{equation}\label{c1}
\psi_{1,0}(\tau) =  \ep^{\frac{\pi^2 \ell}{2G\beta}} \qty(\cos \qty(4\pi^2 /\beta) - 1) . 
\end{equation}

Near $\beta= 2\pi$, the $\psi_{1,0}$ term has the largest exponential, as terms with $c^2-d^2 <0$ have a negative exponential as for $\psi_{0,1}$, while the remaining terms with $c^2 - d^2 >0$ actually satisfy a stronger bound $c^2 - d^2 >1$, and consequently have a smaller exponent.\footnote{This is in contrast to the AdS case, where the terms $Z_{1,0}(\tau)$ and $Z_{0,1}(\tau)$ in the Poincar\'e sum correspond to contributions from the BTZ and the thermal AdS geometry. They undergo a Hawking-Page phase transition at $\beta = 2\pi$, as the saddles exchange dominance. In the dS case, there is no such transition.} 

Thus we have 
\begin{equation}\label{hhdom}
 Z(\beta)= \Psi_\HH(\beta) \approx \frac{1}{\sqrt{\Im \tau} \abs{\eta(\tau)}^2} \psi_{1,0}(\tau) = \cK(\beta) \, \ep^{\frac{\pi^2 \ell}{2G\beta}}, 
\end{equation}
where we identify the thermal partition function in the static patch with the analytically continued boundary wavefunction, and approximate the latter by the $\Psi_{1,0}$ contribution. We have gathered the corrections to the leading exponential behavior in 
\begin{equation}\label{kbeta}
    \cK(\beta) = \frac{(\cos (4\pi^2 /\beta) - 1)}{\cD(\beta)}, \qquad \cD(\beta) \defeq \sqrt{\ic \frac{\beta}{2\pi}} \abs{\eta(\tau)}^2.
\end{equation}
These are one-loop corrections due to boundary gravitons in the wavefunction perspective. 

We can proceed to calculate the entropy by \eqref{entropy}, which gives 
\begin{equation}
    S_{\dS} = \frac{\pi^2 l}{G \beta} + S_{\mathcal K}, 
\end{equation}
reproducing the de Sitter entropy when we set $\beta = 2\pi$, up to the one-loop corrections coming from differentiating $\mathcal K(\beta)$. We take this as evidence that it is necessary to include the full Poincar\'e sum in $\Psi_\HH$ in relating it to the trace over the static patch Hilbert space, as the de Sitter entropy we would expect to find for the static patch comes from the wavefunction perspective from a non-trivial term in the Poincar\'e sum. 

However, there are issues with the one-loop corrections in this calculation, as both the numerator and the denominator in \eqref{kbeta} vanish at $\beta=2\pi$.\footnote{The denominator vanishes as $\eta(\tau) = q^{\frac{1}{24}} \prod_{n=1}^\infty (1-q^n)$, and we have $q = \ep^{2\pi \ic \tau} = \ep^{-T} = \ep^{-\ic\beta} =1$.} The vanishing of the denominator is an overall divergence in the wavefunction, associated with the one-loop determinant of boundary gravitons. From the boundary perspective, this divergence appears physical: when we analytically continue $\Im \tau$, the boundary torus becomes Lorentzian, and when we set $\beta = 2\pi$, this becomes a square Lorentzian torus. It thus has identifications along lightlike directions, and it seems plausible that these lead to a divergent one-loop determinant for the boundary gravitons. 

This will give a formally divergent contribution to the entropy, which we need to regulate. But at least this is an overall divergence. The vanishing in the numerator in \eqref{kbeta} is more concerning, as this is special to the $\psi_{1,0}$ contribution; the corresponding factor in $\psi_{c,d}$ in \eqref{psicdbeta} does not vanish when $\beta=2\pi$. This then implies that if we go close enough to $\beta=2\pi$, the approximation that $\psi_{1,0}$ dominates the wavefunction in \eqref{hhdom} breaks down. 

One way to address these issues is to note that the notion of a static patch relies implicitly on the existence of an observer at $r=0$. Our discussion is in pure gravity, so we have not introduced any notion of an observer, but if we introduce an observer, their back-reaction will slightly modify the geometry, changing the temperature of the cosmological horizon away from $\beta = 2\pi$. Unless the change in $\beta$ is exponentially small in $1/G$, it is sufficient to ensure that the argument that $\psi_{1,0}$ dominates over the other contributions in the Poincar\'e sum is valid. 

Finally, we note that analogous to our relation between the late-time and static patch observables, in AdS black holes, WdW states defined on spacelike slices near the singularity can be related to boundary partition functions \cite{Blacker:2023oan}.

\section{Spectral representation}\label{spectral}
We have identified the trace over the static patch Hilbert space with the boundary torus wavefunction. This offers a new perspective on recent results of Godet \cite{Godet:2024ich} (see also \cite{Godet:2025bju}), who obtained a novel expression for the wavefunction as a $q$-expansion with integer coefficients. His formula is 
\begin{equation}
    \Psi_\HH = \frac{1}{\sqrt{y}\, \abs{\eta(\tau)}^2}  ( \psi[Q] + \psi[\tilde{Q}] - \chi[b]),
\end{equation}
where $c_{\mathrm{dS}} = 1-6Q^2  = 25 - 6 \tilde{Q}^2 = 13 - 6 b^2 - 6 b^{-2}$,
\begin{equation}\label{psiq}
    \psi[Q] = \sqrt{y} \Bigg[ \abs{q}^{\frac{Q^2}{2}} + \sum_{m,n=1}^\infty \mu(m) \, \abs{q}^\frac{n^2}{2m^2 Q^2} + \sum_{m=1}^\infty \mu(m) \sum_{a,d=1}^\infty \qty(q^{h^+_{m,a,d}} \bar q^{h^-_{m,a,d}} + \bar q^{h^+_{m,a,d}} q^{h^-_{m,a,d}}) \Bigg],
\end{equation}
and
\begin{equation}\label{chib}
\begin{split}
    \chi[b] &= \sqrt{y} \Bigg[ \sum_{m=1}^\infty \mu(m) \, \qty(\abs{q}^\frac{b^2}{2m^2} + \abs{q}^\frac{1}{2 b^2 m^2}) \\
    &\qquad+ \sum_{m,l=1}^\infty \mu(m) \mu(l) \sum_{a,d=1}^\infty \qty(q^{h^+_{m,a,l,d}} \bar q^{h^-_{m,a,l,d}} + \bar q^{h^+_{m,a,l,d}} q^{h^-_{m,a,l,d}}) \Bigg],
\end{split}
\end{equation}
where $\mu(m)$ is the M\"obius function, $\mu(m)=0$ if $m$ has an prime factor more than once and $\mu(m)=(-1)^k$ otherwise, where $k$ is the number of distinct prime factors in $m$, and 
\begin{equation}
    h^\pm_{m,a,d} = \frac{1}{4} \qty( amQ \pm \frac{d}{mQ} )^2, \qquad h^\pm_{m,a,l,d} = \frac{1}{4} \qty( a \frac{l b}{m} \pm d \frac{m}{l b})^2. 
\end{equation}

This looks like a sum over a spectrum of states. In particular, for a square torus, we have $q = \ep^{-T}$, and with
$$c_{\mathrm{dS}} = \frac{3\ic \ell}{2G} + 13,$$
we have
$$b^2 = \ic \lambda, \qquad \text{where} \qquad \lambda - \lambda^{-1} = -\frac{\ell}{4G},$$
and the terms in $\chi[b]$ have the form $\ep^{\ic E_m T}$, $\ep^{\ic E_{m,a,l,d} T}$ for real $E$. In the terms in $\psi[Q]$, $E$ has a small imaginary part. Thus, in our proposal, where this wavefunction is identified with the trace over the static patch Hilbert space, we interpret this as giving us a microscopic spectrum associated with the de Sitter static patch. 

There are a number of puzzles worth noting. First, we can see that although the coefficients in the $q$-expansion are integers, they can be negative, as the M\"obius function takes both positive and negative values. Second, in the second term in $\psi[Q]$, there are infinitely many terms in the sum that contribute to a given spectral level, that is, for a given value of $n^2/m^2$, and the sum over $n,m$ with fixed $n^2/m^2$ is divergent. Regulating this divergence is a non-trivial problem mathematically. 

Furthermore, there are some questions about the derivation of this formula. This formula was obtained in \cite{Godet:2024ich} by using a spectral representation of the wavefunction in terms of the Eisenstein series. In the next subsection, we review the derivation and comment on some subtleties in the argument. We then note that there are also terms in $\chi[b]$ coming from Maas cusp forms, and comment on their contribution to the spectrum.

\subsection{Comments on the derivation}
In \cite{Godet:2024ich}, the main idea is to rewrite the wavefunction in a spectral representation. Because of the Poincar\'e sum, the wavefunction \eqref{eq:psi} is a modular-invariant function of $\tau$, so it can be expressed in terms of a basis of such functions. In fact, each of $\psi[Q]$, $\psi[\tilde{Q}]$ and $\chi[b]$ are individually modular-invariant. In \cite{Godet:2024ich}, these were then expressed as a combination of Eisenstein series, 
\begin{equation}\label{psiqspec}
\psi[Q] = \frac{1}{4\pi} \int  \diff  \nu \, \pi^{\ic \nu} \Gamma(-\ic \nu) \qty[Q^{2\ic \nu} + \frac{\zeta(-2\ic \nu)}{\zeta(2\ic\nu)} Q^{-2\ic \nu} ] E_{\frac{1}{2} + \ic \nu}(\tau), 
\end{equation}
\begin{equation}\label{chibspec}
     \chi[b] =  \frac{1}{4\pi} \int  \diff \nu \, \pi^{\ic\nu} \Gamma(-\ic\nu) \frac{2}{\zeta(2\ic \nu)} (b^{2\ic \nu} + b^{-2\ic \nu} ) E_{\frac{1}{2} + \ic \nu}(\tau), 
\end{equation}
where we take the contours to lie just above the real axis. We will comment in the next subsection that this representation can be incomplete, as there can also be contributions from Maas cusp forms. Here we want to simply consider the above integrals and look at how one gets from these to the expressions in \eqref{psiq} and \eqref{chib}. 

The basic idea is to evaluate these integrals by closing the contour in the complex $\nu$ plane. The expressions above are obtained by closing the contour in the lower-half plane, while \cite{Godet:2024ich} argued that closing the contour in the upper-half plane will reproduce the expressions in \cite{Maloney:2007ud}. To understand the detail, it is useful to focus on the part that is independent of $\Re \tau$, i.e., the first two terms in \eqref{psiq} and the first term in \eqref{chib}. These are obtained in the above spectral representation by replacing $E_{\frac{1}{2} + \ic \nu}(\tau)$ by $2 y^{\frac{1}{2}+\ic \nu}$. 

In the first term in \eqref{psiqspec}, the gamma function vanishes as $\nu \to -\ic \infty$, but diverges as $\nu \to + \ic \infty$, so we can actually only close the contour in the lower-half plane. Doing so picks up poles at the zeros of the Gamma function at $\nu = - \ic n$ for $n\in \bbZ_{\geq0}$. Integrating against $2 y^{\frac{1}{2}+\ic \nu}$, this will reproduce the first term in \eqref{psiq}, which is also present in the expressions in \cite{Maloney:2007ud}. 

In \eqref{chibspec}, the combination $\Gamma(-\ic \nu) / \zeta(2\ic \nu)$ vanishes both as $\nu \to -\ic \infty$ and as $\nu \to +\ic \infty$, so we have a choice of how we close the contour. The expression in \eqref{chibspec} is obtained by closing the contour in the lower-half plane and expanding 
\begin{equation}
    \frac{1}{\zeta(2\ic \nu)}= \sum_{m=1}^\infty \mu(m) \, m^{-2\ic \nu}. 
\end{equation}
We can perform these two operations in either order. 

The subtlety comes when we consider the second term in \eqref{psiqspec}. The integrand vanishes when $\nu \to \ic \infty$, but it diverges when $\nu \to -\ic \infty$, so naively we would say we can only close the contour in the upper half-plane. We also note that if we ignored this issue and closed the contour in the lower half plane, the zeros of the zeta function in the numerator would cancel the poles in the gamma function for $\nu = -\ic n$, so there would be no non-trivial contributions to the integral. 

This can be resolved by expanding the zeta function in the numerator as 
\begin{equation}
    \zeta(-2\ic \nu) = \sum_{n=1^\infty} n^{2\ic \nu}. 
\end{equation}
Note that this expansion is only convergent for $\Im \, \nu > \frac{1}{2}$, so to make this expansion we need to shift the contour we are integrating on up slightly, but there doesn't seem to be a problem with this. We then suppose that we can exchange the order of integration and summation to obtain 
\begin{equation} 
\psi^{(0)}[Q] = \sqrt{y} \qty[ \abs{q}^{\frac{Q^2}{2}} + \sum_{m,n=1}^\infty \mu(m)  \frac{1}{2\pi} \int  \diff \nu\, \qty( \frac{\pi n^2 y}{m^2 Q^2} )^{\ic \nu} \Gamma(-\ic\nu) ].
\end{equation}
Now the integrand vanishes as $\nu \to - \ic \infty$, so in each term we can close the contour in the lower half plane. Performing the integral by summing over the poles of the gamma function then gives us the sum in the second term in \eqref{psiq} as desired. This interchange of order of integration and summation is a subtle step since we can't close the contour in the lower half plane if we do the sum first, but this is legitimate when we first integrate term by term.\footnote{In the final expression, it is tempting to resolve the divergences in the double sum in the second term in \eqref{psiq} by expanding the exponential and interchanging the order of summation to write $$\sum_{m,n} \mu(m) \, \abs{q}^{\frac{n^2}{2m^2 Q^2}} = \sum_{k=0}^\infty \frac{1}{k!} \sum_{m,n} \mu(m)\qty( \frac{\pi n^2 \Im \tau}{m^2 Q^2} )^k =  \sum_{k=0}^\infty \frac{1}{k!} \frac{\zeta(-2k)}{\zeta(2k)} \qty( \frac{\pi \Im \tau}{ Q^2} )^k  = 1,$$ where in the last step we used that $\zeta(-2k)=0$ for $k \in \bbZ_{>0}$. But this amounts to re-exchanging the order of integration and summation; that is, this is the answer we would have obtained had we simply ignored the fact that the integrand diverges in the lower half plane and closed the contour there. So we do not believe this manipulation is valid.} Since we take the formula \eqref{psiq} as evidence for our proposal that the wavefunction gives a microscopic definition of $\Tr(\ep^{-\ic HT})$, we want to adopt this order of summation and integration; it would be very useful to have a deeper understanding of this issue.

\subsection{Contribution from Maass cusp forms}
In addition to the expression involving the Eisenstein series, we also have a contribution to the spectrum from the Maass cusp forms. To extract this, we calculate the overlap of the cusp forms with different components of the wavefunction. 

We evaluate the Petersson inner product, which determines the overlap. We use a notation where barred quantities denote the complex conjugates. For the term $\psi[Q]$, the overlap with the cusp forms vanishes for the given range of $x$ with a constant integrand (see \cite{Benjamin:2021ygh}):
\begin{equation}
    (\psi[Q],\nu_n) = \int_{\mathcal{F}} \frac{\diff x \, \diff y}{y^2} \, \psi[Q] \bar \nu_n = \int_0^\infty \frac{\diff y}{y^2} \, \sqrt{y} \, \ep^{-\pi Q^2 y} \int_{-\frac{1}{2}}^{\frac{1}{2}} \diff x \, \bar \nu_n  =0,  
\end{equation}
Here in the second step, we have used the unfolding trick to express the integral over the fundamental domain in terms of the integrals over $x$ and $y$. In contrast, this overlap is non-zero for $\chi[b]$,
\begin{equation}
\begin{split}
    (\chi[b],\nu_n) &= \int_{\mathcal{F}} \frac{\diff x \, \diff y}{y^2} \chi[b] \bar \nu_n = \int_0^\infty \frac{\diff y}{y^2} \, 2 \sqrt{y} \, \ep^{-\pi y(b^2 + b^{-2})} \int_{-\frac{1}{2}}^{\frac{1}{2}} \diff x \, \cos 2\pi x \, \bar \nu_n \\
    &= \int_0^\infty \frac{\diff y}{y}  \ep^{-\pi y(b^2 + b^{-2})} K_{-\ic R_n}(2 \pi y) = \frac{\pi}{2} \frac{\qty(b^{2\ic R_n} + b^{-2\ic R_n})}{R_n \sinh (\pi R_n)},
\end{split}
\end{equation}
where we have used the following decomposition for the Maass cusp forms in terms of Bessel functions:
\begin{equation}\label{maasform}
    \nu_n = \sum_{j=1}^\infty a_j^{(n)} \cos(2\pi j x) \sqrt{y} \, K_{\ic R_n} (2 \pi j y), 
\end{equation}
and used the normalization that $a_1^{(n)} =1$ as in \cite{Benjamin:2021ygh}, and in the last step we used equation (3.65) from \cite{Godet:2024ich}. 

Thus, in the expression for $\chi[b]$, in addition to (3.83) in \cite{Godet:2024ich}, we should also have the contributions from Maass cusp forms:
\begin{equation}\label{chiMaass}
    \chi[b]_{\textrm{Maass}}  = \sum_{n=1}^\infty \frac{\pi}{2} \frac{\qty( b^{2\ic R_n} + b^{-2\ic R_n})}{R_n \sinh (\pi R_n) (\nu_n,\nu_n)} \nu_n(\tau). 
\end{equation}
Plugging in the values for the first five even Maass forms from table 1 in \cite{Benjamin:2021ygh}, the factor in the denominator is of order one.

In equation (3.83) in \cite{Godet:2024ich}, the terms are all of the form $\sqrt{y} \cos(2\pi j x) \, \ep^{\ic y E}$ for real $E$ (as $b^2$ is imaginary). We would like to write this additional contribution in a similar form. We write the contribution using the following representation of the Bessel function:
\begin{equation}
    K_{\ic \nu}(2\pi y) = \frac{1}{2} \int \frac{\diff t}{t} \, \qty( \frac{\pi y}{t} )^{\ic\nu} \mathrm{exp}\qty( -t -\frac{\pi^2 y^2}{t} )= \frac{1}{2} \int  \diff \tau \, \ep^{-\ic \nu \tau} \ep^{-2\pi y \cosh \tau}, 
\end{equation}
where we have made a change of variable $t = \pi y \ep^\tau$. Then equation \eqref{chiMaass} has the following integral representation:
\begin{equation}
\begin{split}
    \chi[b]_{\textrm{Maass}}  = \sqrt{y} \sum_{j=1}^\infty \cos (2\pi j x) \int_{\mathbb R} \diff \tau \, \ep^{-2\pi y \cosh \tau} \sum_{n=1}^\infty &a_j^{(n)}\frac{\pi}{2} \frac{\qty( b^{2\ic R_n} + b^{-2\ic R_n}) \ep^{-\ic R_n \tau}}{R_n \sinh (\pi R_n) (\nu_n,\nu_n)}. 
\end{split}
\end{equation}
To obtain the desired form; we shift the contour of integration in $\tau$ to $\tau = \ic \frac{\pi}{2} + \tilde{\tau}$, and write $b^2 = \ic \lambda$. This gives us:
\begin{equation}
\begin{split}
    \chi[b]_{\textrm{Maass}}  = \sqrt{y} \sum_{j=1}^\infty \cos (2\pi j x) \int_{\mathbb R} \diff \tilde{\tau} \, \ep^{-2\pi \ic y \sinh \tilde{\tau}} \sum_{n=1}^\infty &a_j^{(n)}\frac{\pi}{2} \frac{\qty(\lambda^{\ic R_n} + \qty(-\frac{1}{\lambda})^{\ic R_n}) \ep^{-\ic R_n \tilde{\tau}} }{R_n \sinh (\pi R_n) (\nu_n,\nu_n)},
\end{split}
\end{equation}
which is of the form $\sqrt{y} \cos(2\pi j x) \, \ep^{\ic y E}$, with an integral over $E$. Thus, we can represent the contribution of the Maass cusp forms to $\chi[b]$ using a spectral representation,
\begin{equation}
    \chi[b]_{\textrm{Maass}}  = \sqrt{y} \sum_{j=1}^\infty \cos (2\pi j x) \int_{\mathbb R}  \diff \tilde{\tau} \ \ep^{-2\pi \ic y \sinh \tilde{\tau}} \rho_j(\tilde{\tau}).
\end{equation}
Here, the spectral density takes the following form:
\begin{equation}
    \rho_j(\tilde{\tau}) = \sum_{n=1}^\infty a_j^{(n)}\frac{\pi}{2} \frac{\qty(\lambda^{\ic R_n} + \qty( -\frac{1}{\lambda})^{\ic R_n}) \ep^{-\ic R_n \tilde{\tau}}}{R_n \sinh (\pi R_n) (\nu_n,\nu_n)}. 
\end{equation}

The difference here from the Eisenstein series spectrum is that, because the Maass forms involve a sum rather than an integral, we get a continuous rather than a discrete set of terms in the spectrum. Interpreting this additional contribution in terms of $\Tr(\ep^{-\ic HT})$ is thus less clear. As a first step, we would like to better understand the form of $\rho_j(\tilde{\tau})$.

\section{Discussion}
We have proposed a new Lorentzian path integral approach to studying quantum gravity in de Sitter space, inspired by the relation between the spectral form factor and double cone geometries in AdS \cite{Saad:2018bqo}. We view the Lorentzian static patch geometry \eqref{staticp} with the time direction periodically identified, $t \sim t + T$, as an approximation to $\Tr(\ep^{-\ic HT})$, the trace over a putative Hilbert space associated to the static patch. Resolving the singularity in the geometry at $r=\ell$ by a complex deformation of the metric, we have related this to the wavefunction on $\cI^+$. 

This saddle is just one of an infinite family of contributions to the wavefunction \eqref{HH1}, related by modular transformations of the boundary torus at $\cI^+$. Our main conjecture is then to use the full wavefunction $\Psi_\HH$ to define what we mean by $\Tr(\ep^{-\ic HT})$. 

We have given two pieces of evidence for this conjecture.  First, we showed that when we analytically continue to imaginary values, $T \to \ic \beta$, the wavefunction is dominated by another saddle $\Psi_{1,0}$ related by an S-transformation and that the contribution of this saddle correctly reproduces the de Sitter entropy. Second, the calculation of $\Psi_\HH$ in \cite{Godet:2024ich} shows that at least the scalar part of the wavefunction can be written as a sum over a set of states labeled by discrete parameters with integer degeneracies, supporting the idea that this does correspond to a trace over a set of states. (Although the Maas cusp form contributions for the non-scalar parts produce a continuous contribution, somewhat reducing the force of this argument.)

There are a number of remaining mysteries in this story. First, the fundamental basis of our argument is the resolution of the singularity in the periodically identified static patch at the horizon by complexifying the geometry. This is inspired by work on the double cone in AdS \cite{Saad:2018bqo, Chakravarty:2024bna}, but unlike in those contexts, our complex saddle does not satisfy the KSW condition \cite{Kontsevich:2021dmb, Witten:2021nzp}. We have noted that this condition can also be violated in other cases, including higher-dimensional versions of the double cone \cite{Chakravarty:2024bna}. Understanding the scope of the KSW condition and whether the complex saddles we have used here are actually allowable is clearly of central importance to validating our ideas. 

Another important direction involves the de Sitter entropy. In our calculation, the usual Bekenstein-Hawking entropy is correctly reproduced by the classical action of the S-transformed saddle $\psi_{1,0}$, but there are one-loop divergences when we set the temperature equal to the de Sitter temperature.  It is clearly important to understand the physics of these divergences. We have proposed that the divergences could be resolved if we considered a theory including an observer at the origin of the static patch. Introducing an observer would also make it possible to make contact with the interesting recent work on de Sitter entropy from an algebraic point of view \cite{Chandrasekaran:2022cip}, which would be a promising direction for future work. We have also suggested that the overall divergence could be related to the fact that the analytically continued boundary torus has a compact null direction -- it would be interesting to flesh out this suggestion. 

In the discussion of the spectrum, there are also a number of questions for the future. We would like to understand the meaning of the negative contributions to the trace better. Also, while the part of the energy which goes like $1/G$ is real, some of the terms have an order one imaginary part to the energy, whose physical meaning is also not yet clear. 

There are also some more mathematical issues: we would like to understand how to treat the divergences from the sum over $m,n$ for given $m/n$ in the second term in \eqref{psiq}. In the non-scalar part of the wavefunction, there is also a contribution from the Mass cusp forms, which gives a continuous contribution to the would-be density of states. Characterizing this contribution more fully is a significant challenge for the future. 

Our focus here has been on pure gravity in three dimensions. It would be interesting to attempt to extend these ideas to higher dimensions. The basic idea of relating the trace over the static patch to some wavefunction on $\cI^+$ could carry over, although we have less control of the nature of that wavefunction in higher dimensions.

\section*{Acknowledgements}
We thank Victor Godet and Viraj Meruliya for related discussions. The work of J.C. is supported by the National Science and Engineering Council of Canada (NSERC) and the Canada Research Chair program, reference number CRC-2022-00421. The work of S.F.R. is supported in part by STFC through grant ST/T000708/1. A.M. and K.N. are supported in part by the Natural Sciences and Engineering Research Council of Canada (NSERC), funding reference number SAPIN/00047.

\appendix
\section{\texorpdfstring{Gravity in AdS$_3$ and MWK partition function}{Brief review of AdS pure gravity}}
Einstein gravity in three dimensions has no local degrees of freedom. With a negative cosmological constant, the spacetime is locally AdS. However, we still have non-trivial geometries that take the form AdS$_3 / \Gamma$, where $\Gamma \subset \SO(2,2)$. An example of this is BTZ which has an identification on the angle $\phi \sim \phi + 2\pi$. 

The primary states in the theory represent genuine particles or black holes in AdS. Although there are no local gravitons, we still have boundary gravitons. There are essentially \textit{edge modes} associated with large diffeomorphisms. These are represented by descendant states of the following form:
\begin{equation}
{\rm Boundary \, \, Gravitons:} \qquad \ket{\psi_{{n_k},{k}}} =  \{ L_{-1}^{n_1} \dots L_{-k}^{n_k} \} \ket{0},
\end{equation}
where $L$'s denote Virasoro generators, and where $\ket{0}$ denotes the $\SL(2,\bbR)$ invariant vacuum. The action of each $L$ is essentially a boundary diffeomorphism, and hence a general descendant state is dressed with boundary gravitons. 

The BTZ geometry is related to the thermal AdS geometry by a modular transformation on the boundary torus, and by a large diffeomorphism in the bulk. More generally, we could do any $\SL(2,\bbZ)$ modular transformation to obtain a new geometry. However, the action of the matrix $T$ given by
\begin{equation}
T = \begin{pmatrix}
    1 & 1\\
    0&1
\end{pmatrix}
\end{equation}
does not give a new bulk geometry. This implies that there are an infinite number of geometries labeled by $\SL(2,\bbZ)/\bbZ$, where the quotient is essentially the contribution $T^n$.

Each non-trivial geometry is characterized by a choice of cycle on the torus that is contractible in the bulk. In general, the elements of $\SL(2,\bbZ)/\bbZ$ are uniquely parametrized by the co-prime pair $(c,d)$, where the cycle given by $c t + d \phi$ is contractible in the bulk. 

Let us now evaluate the thermal AdS partition function.
\begin{equation}
{Z}_{0,1}(x,y) ={\rm Tr}_\cH \, \ep^{-\beta H + \ic \theta J}
\end{equation}
The Hilbert space $\cH$ receives contributions from the vacuum sector. This includes the SL$(2, R)$ invariant vacuum and boundary graviton states, which receive non-zero contributions. The thermal partition function includes contributions from all possible states apart from $L_{-1}$, since $L_{-1}$ annihilates the vacuum. Define $q = \ep^{2 \pi \ic \tau}$, where 
\begin{equation}
\tau = \frac{1}{2\pi} \qty(\theta + \ic \beta )
\end{equation}
Using this, we now obtain the expression for the thermal partition function:
\begin{equation}\label{zt01}
{Z}_{0,1}(\tau) ={\rm Tr}_\cH q^{L_0} \bar{q}^{\bar{L}_0} = q^{-\frac{c}{12}} \bar{q}^{-\frac{c}{12}} \prod_{n=2}^\infty \frac{1}{\abs{1-q^n}^2}
\end{equation}
The powers of $q$ outside the product denote the contribution from the ground state, while the product is the contribution from the boundary gravitons.

\subsection*{MWK partition function and Poincar\'e sum}\label{mwkpf}
 For pure gravity in AdS, using the Dedekind $\eta$ function, the MWK partition function takes the form of a Poincar\'e sum over the thermal AdS partition function and its modular images under SL$(2,\bbZ)/\bbZ$:
\begin{equation} 
    Z_{\rm MWK}(\tau) =  Z_{0,1}(\tau) + \sum_{c,d} Z_{c,d}(\tau)
\end{equation}
In \eqref{mwk}, the first term is 
\begin{equation}\label{mwk1}
Z_{0,1}(\tau) = \frac{1}{\abs{\eta(\tau)}^2} \abs{q}^{-\frac{c-1}{12}} \abs{1-q}^2,
\end{equation}
where $q = \ep^{2\pi \ic \tau}$. The other terms in the Poincar\'e sum $Z_{c,d}(\tau)$ denote its modular images under the action of $\SL(2,\bbZ)/\bbZ$  which are given by 
\begin{equation}
Z_{c,d}(\tau) = Z_{0,1}(\gamma \tau),\qquad\gamma \tau = \frac{a \tau+b}{c\tau + d}, \quad ad-bc =1,
\end{equation}
such that the sum is over relatively prime $(c,d)$ with $c>0$. We write $Z_{0,1}(\tau) $ as follows:
\begin{equation} 
Z_{0,1}(\tau) = \frac{1}{\abs{\eta(\tau)}^2} \ep^{2\pi \frac{(c-13)}{12}y} (\cosh 2\pi y - \cos 2\pi x). 
\end{equation}
We are interested in the behavior of the MWK partition function for zero twist $x=0$ evaluated at $y=1$. To see this, we define a new parameter:
\begin{equation}\label{mwkpar}
y' = \Im \gamma \tau = \frac{y}{(cx+d)^2 + c^2 y^2},  
\end{equation}
so near $y=1$, $y' = 1/\qty( d^2+c^2) <1$ except for the term with $c=1, d=0$. Thus, with a large central charge, the other terms in the Poincar\'e sum are exponentially suppressed (although the whole sum diverges, we are interested in the relative suppression). There is an exchange of dominance at $x=0, y=1$ between the two saddles $Z_{0,1}(\tau)$ and $Z_{1,0}(\tau)$, which is the familiar Hawking-Page phase transition between thermal AdS at temperature $\beta$ and BTZ with temperature $1/\beta$.

\section{Wavefunction normalization}\label{norm}
We provide some arguments regarding the normalization of the wavefunction $\Psi_\HH$. Recall that the MWK partition function is evaluated over an $\SL(2,\bbR)$ invariant vacuum, which is invariant under global conformal transformations using Virasoro generators $L_0$ and $L_{\pm 1}$. Upon analytic continuation, these conformal Killing vectors on the torus slicings lead to zero modes in the de Sitter path integral. 

Our goal here is to understand the origin of these zero modes from the gravity side. From the perspective of the Wheeler-DeWitt equation at late time dS, the wavefunctional $\Psi_\HH$ on a late-time torus slicing is a spatial diffeomorphism-invariant functional that obeys Weyl transformation properties \cite{Chakraborty:2023yed, Godet:2024ich}.\footnote{See \cite{Freidel:2008sh} for an analogous Weyl factor that arises in the context of the radial WdW equation.} The global conformal transformations on the late-time torus correspond to the residual gauge transformations $\SL(2,\bbR)$ upon fixing Weyl and diffeomorphisms. We find that fixing them appropriately leads to the correct normalization of the wavefunction.

Thus, for a late-time observable involving the state $\Psi_\HH$, we need to fix the gauge symmetries within the path integral over the space of metrics, by dividing with the volume of the gauge group \cite{Chakraborty:2023los}:
\begin{equation}
    \expval{A}{\Psi_\HH} = \frac{\mathcal{N}^2}{\mathrm{Vol} \qty( \mathrm{diff} \times \mathrm{Weyl} )}\int { \diff g} \, \abs{\Psi(g)}^2 A(g)
\end{equation}
where $A$ denotes a string of operator insertions on the late-time slice, and where $g$ denotes the metric on the late-time Cauchy slice. 

Fixing the diffeomorphism and Weyl transformations leads to residual global conformal transformations $\SL(2,\bbR)$. Performing the division directly is difficult since $\SL(2,\bbR)$ is a non-compact group, which gives rise to an infinity. This requires the familiar procedure of fixing three points, analogous to string theory \cite{Polchinski:1998rq}. 

A more powerful procedure to implement the diffeomorphism and Weyl gauge-fixing is by using BRST quantization. We schematically list the key steps here.\footnote{In our present case, we utilize the derivation of BRST quantization of bosonic string as in \cite{Polchinski:1998rq}. Our treatment here is based on Appendix C of \cite{Chakraborty:2023los}, where a similar BRST quantization of gravity coupled to matter was performed in late-time dS$_{d+1}$ over the space of WdW solutions, where we encounter diff $\times$ Weyl symmetries. See also \cite{Witten:2022xxp} for Weyl transformations in the context of WdW in AdS and the BRST operator.} We consider the expectation value of the observable $A$ given by:
\begin{equation}
\expval{A}{\Psi_\HH} = \mathcal{N}^2 \int { \diff g} \, \abs{\Psi(g)}^2 A(g) Z_{\rm add} 
\end{equation}
Here the additional term in the action $Z_{\rm add} $ is given by the following expression:
\begin{equation} \label{zadd}
Z_{\rm add} = \int { \diff c} \, { \diff \bar{c}} \, { \diff b} \,{ \diff \bar{b}} \, { \diff B} \,{ \diff B_i} \, \ep^{-S_{\rm gh}-S_{\rm g.f.}}
\end{equation}
where $b$ and $\bar{b}$ denote the Weyl ghosts, $c$ and $\bar{c}$ denote the diffeomorphism ghosts, while $B_i$ and $B$ denote the Nakanishi-Lautrup fields which implement fixing the diffeomorphism and Weyl invariance, respectively. Integrating out the $b$ fields leads us back to the answer obtained by fixing the three points. The gauge fixing part in the BRST action takes the following form:
\begin{equation}
S_{\rm g.f.} = \ic \int \diff^2 z \, \qty( B F_W(g) + B_i F^i_D(g) ),
\end{equation}
where $F_W(g)$ and $F^i_D(g)$ denote the Weyl and diffeomorphism fixing conditions respectively. The term $S_{\rm g.f.}$ denotes the gauge fixing action using the Nakanishi-Lautrup fields, while the $S_{\rm gh}$ denotes the ghost action. In three dimensions, integrating out the Nakanishi-Lautrup fields essentially leads to a bc ghost action on the late-time slice \cite{Polchinski:1998rq}.
\begin{equation}
    S_{\rm bc} = \frac{1}{2\pi} \int \diff^2z \, \qty(\bar{b} {\del} \bar{c} + b \bar{\del} c )
\end{equation}
The bc ghost system has central charge $c_{bc}=-26$. Therefore to ensure that the associated BRST charge $Q_B$ is nilpotent for pure three dimensional gravity, i.e. $\{ Q_B , Q_B \} =0$, we require the familiar one-loop contribution of $13$ in $c_\dS=\frac{3\ic \ell}{2G} + 13$. Additionally, bc ghost system gives us a factor $Z_{bc}(\tau) = y \abs{\eta(\tau)}^4$, which fixes the overall normalization as follows \cite{Godet:2024ich}.
\begin{equation}
\mathcal{N}^2 = \frac{1}{y \abs{\eta(\tau)}^4}
\end{equation}
This gives our desired normalization for the Hartle-Hawking state. This also implies that the allowed observable at late time, i.e., the cosmological correlator with the late-time insertion $A$ is defined as the following expression:
\begin{equation}
\frac{\expval{A}{\Psi_\HH}}{\braket{\Psi_\HH}} = \int { \diff g} \, \abs{\Psi(g)}^2 A(g) Z_{\rm add} 
\end{equation}
where $Z_{\rm add}$ is defined in \eqref{zadd}. The above procedure is reflected while performing the analytic continuation: the global conformal transformations on the late-time torus correspond to the residual gauge transformations $\SL(2,\bbR)$ upon fixing Weyl and diffeomorphisms. 

Finally, another simple way to understand the normalization of the wavefunction is as follows: the Wheeler-DeWitt equation is basically the Klein-Gordon equation of a particle \cite{Carlip:1993ak, Carlip:1995zj}. Thus, the norm of the wavefunction can be understood as the Klein-Gordon norm over its solution space \cite{Godet:2024ich}.

\bibliographystyle{JHEP}
\bibliography{refs}
\end{document}